\documentclass[12pt]{article}
\usepackage{latexsym,graphicx,multirow}
\usepackage{float}
\usepackage{amssymb}
\usepackage{amsmath}
\usepackage{amscd}
\usepackage{amsthm}
\usepackage[left=2cm,top=2.5cm,right=2.5cm,bottom=1.5cm]{geometry}
\usepackage{hyperref}
\usepackage{epstopdf}
\usepackage{cite}
\usepackage{caption}
\usepackage{subcaption}
\usepackage[utf8]{inputenc}
\usepackage[T1]{fontenc}
\begin{document}
	
\begin{center}
\large{\bf{ Cosmographic analysis of a closed bouncing universe with the varying cosmological constant in $f(R,T)$ gravity}} \\
		\vspace{10mm}
\normalsize{Vinod Kumar Bhardwaj $^{1}$, Anirudh Pradhan$^{2}$, Nasr Ahmed$^{3,4}$, A. A. Shaker $^{4}$ }\\
		\vspace{5mm}
\normalsize{$^{1}$ Department of Mathematics, Institute of Applied Sciences \& Humanities, GLA University, Mathura-281 406, Uttar Pradesh, India } \\
\vspace{2mm}
\normalsize{$^{2}$ Centre for Cosmology, Astrophysics and Space Science (CCASS), GLA University, Mathura-281 406, Uttar Pradesh, India.} \\
\vspace{2mm}
\normalsize{$^{3}$ Mathematics Department, Faculty of Science, Taibah University, Saudi Arabia.} \\
\vspace{2mm}
\normalsize{$^4$ Astronomy Department, National Research Institute of Astronomy and Geophysics, Helwan, Cairo, Egypt }\\
      \vspace{5mm}
$^{1}$dr.vinodbhardwaj@gmail.com \\
	\vspace{2mm}
$^{2}$pradhan.anirudh@gmail.com \\
	\vspace{2mm}
$^{3}$nasr.ahmed@nriag.sci.eg \\
	\vspace{2mm}
$^{4}$shaker@nriag.sci.eg \\		
	\vspace{2mm}


\end{center}
	
\begin{abstract}
	Modeling of matter bounce in $f(R,T)$ gravity has been presented with no violation of the null energy condition. Only a closed universe with
	negative pressure is allowed in good agreement with some recent observations which favor a universe with positive curvature. Our results agree with some recent works in which a combination of positive curvature and vacuum energy leads to non-singular bounces with no violation of the null energy condition. The stability of the model has been discussed. The cosmographic parameters are developed for the derived model to explain the accelerated expansion of the universe.
	
\end{abstract}


\section{Introduction and motivation}
Recent observational predictions that our Universe is going through a phase of accelerated expansion from the Supernova Cosmology Project collaboration \cite{1,2}, Supernova Search Team collaboration \cite{3,4}, WMAP collaboration \cite{5,6}, and Planck Collaboration \cite{7} have opened up new avenues in modern cosmology. These findings suggest that our Universe is dominated by a strange cosmic fluid with huge negative pressure nicknamed as dark energy (DE), which accounts for $\simeq 3/4$ of the critical density (\cite{8,9} for detailed review). Furthermore, studies of the cosmic microwave background (CMB) show that the Universe is flat on enormous scales. Because there isn't enough stuff in the Universe to achieve this flatness-neither ordinary nor dark matter-the discrepancy must be attributed to a dark energy. The acceleration of the expansion of the Universe is caused by the same dark energy. Furthermore, the influence of dark energy appears to change throughout time, with the expansion of the Universe slowing and speeding up. According to the Wilkinson Microwave Anisotropy Probe (WMAP) satellite experiment, dark energy makes up $73\%$ of the Universe's substance, non-baryonic dark matter makes up $23\%$, and ordinary baryonic matter and radiation make up the remaining $4\%$. 

Einstein's general theory of relativity is widely acknowledged as the most successful theory of gravity and the foundation for developing cosmological models of the Universe. The modified theory of gravity acquired favour among cosmologists due to Einstein's theory's incompatibility with Mach's principle and a current scenario of the Universe undergoing a late time rapid expansion. They claim that a modified theory of gravity can better explain the Universe's late-time acceleration. To develop a better modified theory of gravity, researchers have made different adjustments to Einstein's theory. $f(R)$ gravity \cite{10,11}, $f(G)$ gravity \cite{12,13,14,15,16}, $f(T)$ gravity \cite{17,18}, 
$f(R,T)$ gravity \cite{19}, and $f(R,T,R_{\mu\nu}T^{\mu\nu})$ gravity \cite{20,21} are the most well-known modified theories of gravitation. 
Modified gravity theories surely give a method of comprehending the DE problem and the prospect of rebuilding a gravitational field theory capable of explaining the Universe's late-time rapid expansion. By substituting scaler curvature $R$ in Einstein-Hilbert action with an arbitrary function of Ricci scaler $R$, known as $f(R)$ gravity, Nojiri and Odintsov \cite{22} constructed a modified theory. Harko {\it et al.} \cite{19} have presented a $f(R,T)$ extension of $f(R)$ gravity that incorporates the trace $T$ of the energy momentum tensor $T_{ij}$. For numerous types of $f(R,T)$, they have developed the associated field equations in metric formalism. They also claim that, due to the interaction of matter and geometry, the cosmic acceleration in $f(R,T)$ gravity theory comes from both geometrical and matter content contributions. Several writers have looked into the astrophysical and cosmological consequences of $f(R,T)$ gravity extensively \cite{23,24,25,26,27,28,29,30}. 

While the standard Big Bang model is very successful, it suffers from a number of problems among them is the initial singularity. Although the inflationary scenario has provided solutions to some problems, the problem of the initial singularity remained unanswered \cite{31}. The Big Bounce represents an alternative cosmological theory where the initial singularity problem does not exist and the expanding universe originates from a prior contracting phase of minimal size \cite{32,33,34,35,36,37,38,39,40,41,42,43}. Based on this bouncing scenario, the process of contraction-expansion may continue forever. The bouncing scenario has been investigated in the context of several modified gravity theories \cite{36,44,45,46,47,48,49} and teleparallel gravity \cite{50}. Several bouncing cosmological models have been presented in the literature among them is the Matter Bounce Scenario (MBS) which has attracted a special attention \cite{51,52,53,54,55,56,57,58}. According to the MBS, the universe, at a very early epoch, is nearly matter-dominated and then the evolution slowly continues towards a bounce. Because all different cosmic parts are assumed to be in causal 
contact at the bounce, the horizon problem also doesn't appear in this scenario. Then, the cosmic expansion starts and continues in parallel to the Big Bang model. Some open questions about the MBS have been discussed in \cite{37}.

While the flat universe is supported by many observational and theoretical works \cite{59,60,61,62,63}, Some recent observations of CMB anisotropies also suggest a closed universe \cite{64,65,66,67}. The current theoretical work supports the 
closed universe where we found that the existence of a stable bouncing cosmology in $f(R,T)$ gravity is related to the positive curvature. Cosmography is a criteria for determining which model performs better than others when compared to cosmological data \cite{68}. In cosmography, we describe the universe through model independent filling procedure without need of any assumption given a priori on universe's cosmology, just its geometry and flatness. The cosmographic evaluation and usefulness in distinct theoretical models under various circumstances can be seen in literature \cite{69,70,71,72,73}. The following is a description of the paper's structure: In section $2$, a matter-bounce solution to the modified cosmological equations has been provided with a complete analysis for the evolution of different parameters. Section $3$ is dedicated to the study of the stability of the model. In section $4$, cosmographic analysis of the model is described. Section $5$ is the conclusion.
\section{Cosmological equations and solutions} \label{sol}
Given the action of $f(R,T)$ modified gravity \cite{8}
\begin{equation} \label{1}
S=\frac{1}{16\pi}\int{\sqrt{-g} f(R,T)d^{4}x}+\int{\sqrt{-g} L_{m} d^{4}x} , 
\end{equation}
where $ f (R, T) $ is an arbitrary function of Ricci scalar $ R $,  trace of energy momentum tensor $ T_{\mu \nu} $ and $ L_{m} $ is 
the Lagrangian density of matter field.\\
The gravitational field equation of $ f(R, T)  $ gravity on varying with respect to the metric tensor $ g_{\mu \nu} $ is given by
\begin{eqnarray}\label{2}
&F_{1}(R,T) R_{\mu \nu}-\frac{1}{2}f(R,T)g_{\mu \nu}+\left(g_{\mu \nu}\nabla^{\mu}\nabla_{\mu}-\nabla_{\mu}\nabla_{\nu}\right) F_{1}(R,T)\nonumber\\
&=\left[8 \pi  -F_{2}(R,T)\right] T_{\mu \nu}-F_{2}(R,T)\Theta_{\mu \nu}
\end{eqnarray}
where, $ F_{1}(R,T)=\frac{\partial f(R,T)}{\partial R}, F_{2}(R,T)=\frac{\partial f(R,T)}{\partial T}  $ and $ \nabla_{\mu} $ is 
the co-variant derivative. The energy-momentum tensor for a perfect fluid distribution, $ T_{\mu \nu}=-p g_{\mu \nu}+(p+\rho)u_{\mu}u_{\nu} $ 
and $ \Theta_{\mu \nu}=g^{\alpha \beta} \frac{\delta T_{\alpha \beta}}{\delta g^{\mu \nu}} $ are derived from the matter Lagrangian density 
$ L_{m} $. Taking the matter Lagrangian density as $ L_{m}=-p $, gives $ \Theta_{\mu \nu}=-p g_{\mu \nu}-2T_{\mu \nu} $. Here, $ \rho $ and 
$ p $ are the energy density and pressure, respectively.\\
In the present paper, we have taken $ f(R,T)=R+2 f(T) $ as functional form of modified gravity, where $ f(T) $ is the trace of energy-momentum tensor.

The corresponding field equations reduce to
\begin{equation} \label{3}
R_{\mu \nu}-\frac{1}{2}R g_{\mu \nu}=8 \pi T_{\mu \nu}+2 f_{T} T_{\mu \nu}+[f(T)+2p f_{T}]g_{\mu \nu}
\end{equation}  
where $ f_{T} $ is the partial derivative of $ f $ with respect to T. 	
The $ f(R,T) $ modified field equations with varying $ \Lambda $ are given by \cite{74}
\begin{equation} \label{4}
G_{\mu \nu}=\left[8 \pi +2 f_{T}\right] T_{\mu \nu} +[f(T)+2p f_{T}+\Lambda(t)]g_{\mu \nu}
\end{equation} 
Where $G_{\mu\nu}=R_{\mu\nu}-\frac{1}{2} R g_{\mu\nu} $ is the Einstein tensor.
Assuming $ f(T)=\lambda T $, $ \lambda $ is a constant, Eq. (\ref{4}) can be written as
\begin{equation} \label{5}
G_{\mu \nu}=\left[8 \pi +2 \lambda\right] T_{\mu \nu} +[\lambda \rho-\lambda p +\Lambda(t)]g_{\mu \nu}
\end{equation} 
where the time-dependent cosmological constant is considered as  \cite{7,75,76,77,78,79,80}:
\begin{equation}\label{6}
\Lambda(t)=-\frac{1}{2}(\rho-p) ,
\end{equation} 
which relates the cosmological constant to the thermodynamical work density \cite{81} $W=\frac{1}{2}(\rho-p)$ by $\Lambda(t)=-W(t)$ where $\rho$ and $p$ are the energy density and pressure. This gives a thermodynamical interpretation to the varying $\Lambda$ in this specific $f(R,T)$ reconstruction. Another thermodynamical interpretation to the cosmological constant appears in black hole physics as thermodynamic pressure \cite{82,83}. The FRW metric is 
\begin{equation}\label{7}
ds^{2}=-dt^{2}+a^{2}(t)\left[ \frac{dr^{2}}{1-kr^2}+r^2d\theta^2+r^2\sin^2\theta d\phi^2 \right] 
\end{equation} 
Equation (\ref{5}) with the metric (\ref{7}) gives
\begin{eqnarray}
\frac{\dot{a}^2+k}{a^2} &=& \frac{(8\pi+3\lambda)\rho-\lambda p+\Lambda(t)}{3}.  \label{8}\\
\frac{\ddot{a}}{a} &=&  \frac{-4(3\pi+\lambda) p-4 \pi \rho+\Lambda(t)}{3}. \label{9}
\end{eqnarray}
The condition of matter bounce is explained by bouncing cosmology, in which cosmological models describe the universe's transition from earlier cosmic contraction to current accelerating expansion with non-singular bounce \cite{54}. Under the reported spectrum of cosmic fluctuation, these bouncing models propose an alternative to inflation \cite{41,42,43,52,58}. The transmission of variations from the initial contracting phase to the expansion period can be dealt precisely because cosmic bounce is non-singular.\\
General expression of matter bounce scenario can be expressed as $ a(t)=a_{0}\left(M \bar{\rho} t^2+Q\right)^Z$
where $ M, Q, Z $ are constants and $ \bar{\rho}\approx 1.88\times 10^{-29} h^2 gcm^{-3} $ is the critical density of the universe. 
$ M=2/3, 3/2, 3/4 $ or $ 4/3 $, $ Q=1 $ and $ Z=1/3 $ or $ 1/4 $ \cite{45,54,38,39}. On the basis of above motivation, in the present work, we consider the following scale factor for a variant non-singular bounce \cite{48,36}:
\begin{equation} \label{10}
a(t)=\left(A t^2+1\right)^n
\end{equation}
From the above equation, we find that $ \dot{a}<0 $ for $ t<0 $ and $ \dot{a}>0 $ for $ t>0 $; which ensures prior contracting phase and later expanding phase. At $ t=0, a(t)\neq 0 $; which ensures the null value of Hubble parameter at the bounce point.\\
The MBS can be studied via this ansatz when $n=\frac{1}{3}$. The deceleration and Hubble parameters $q$ and $H$ are given as 
\begin{equation} \label{11}
q(t)=-\frac{\ddot{a}a}{\dot{a}^2}=-\frac{(2n-1)A t^2+1}{2n A t^2}  \;\;\;\;\;\;\;,  \;\;\;\;\;\;\; H(t)= \frac{2n A t}{ A t^2+1}
\end{equation}
\begin{figure}[H]
	\centering
	(a)\includegraphics[width=5cm,height=4cm,angle=0]{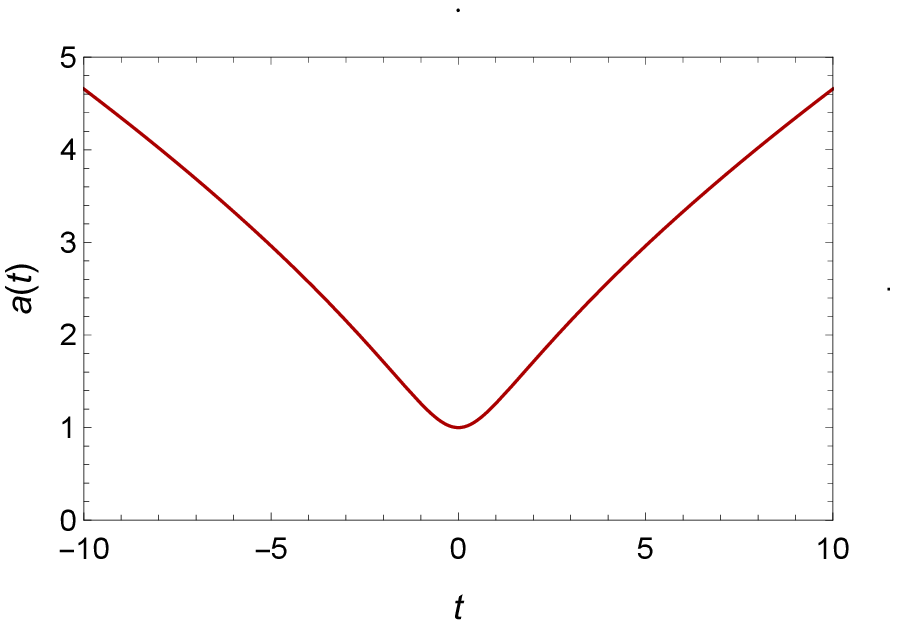}
	(b)\includegraphics[width=5cm,height=4cm,angle=0]{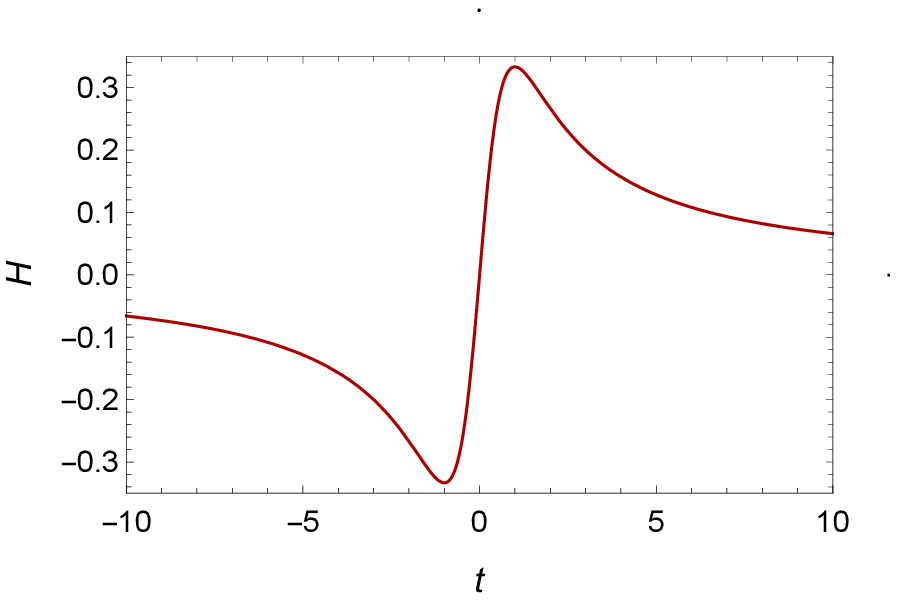}
	(c)\includegraphics[width=5cm,height=4cm,angle=0]{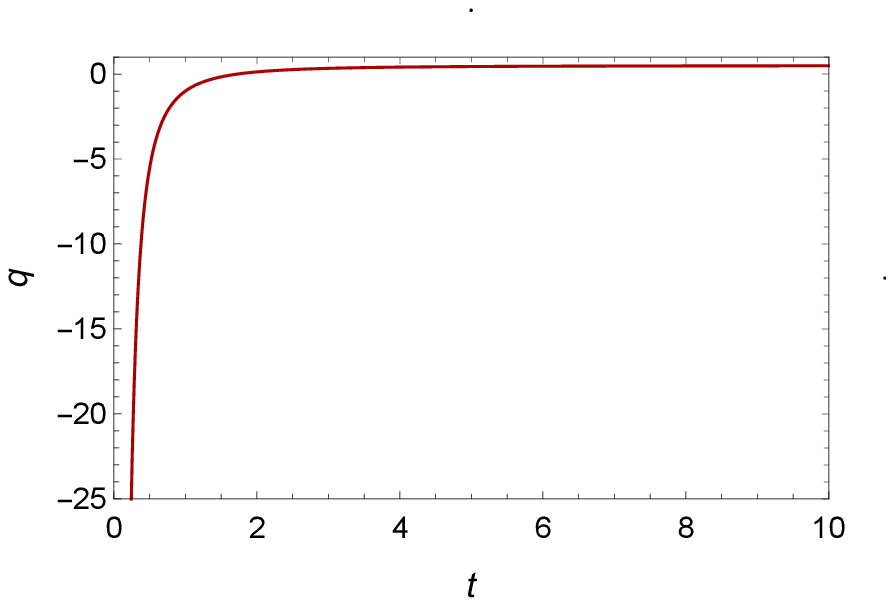}
	\caption{ (a). The scale factor, (b)Hubble parameter \& (c)deceleration for the MBS ($n=\frac{1}{3}$) and $A=1$.}\label{fig:cas5}	
\end{figure}
The pressure and energy density are given as 
\begin{equation}\label{12} 
p(t)=\frac{f(t) g_{1}(t)}{f_1(t)} \;\;\;\;\;\;\; \& \;\;\;\;\;\;\; \rho(t)=\frac{f(t) g(t)}{f_1(t)}
\end{equation}
Where
\begin{eqnarray*}  
	f(t)& = & \left(A t^2+1\right)^{-2 (n+1)}\\
	f_1(t)&=& 2 \big[\lambda (4 \lambda -1)+4 \pi  (6 \lambda -1)+32 \pi ^2\big]\\
	g(t)&=& 2 A n \left(A t^2+1\right)^{2 n} \big[A t^2 \big(2 \lambda +12 (\lambda +4 \pi ) n-1\big)-2 \lambda +1\big]\\
	&+&k (8 \lambda +24 \pi -1) \left(A t^2+1\right)^2\\
	g_1(t)&=& -2 A n \left(A t^2+1\right)^{2 n} \big[A t^2 (-6 \lambda +12 \lambda  n+1)+16 \pi  \left(A (3 n-1) t^2+1\right)+6 \lambda -1\big]\\
	&-&(1+8 \pi ) k \left(A t^2+1\right)^2\\
\end{eqnarray*}  
So, the cosmological constant (\ref{6}) now becomes
\begin{equation}\label{13}   
\Lambda(t)=-\frac{2 \left(A t^2+1\right)^{-2 (n+1)} \left(k \left(A t^2+1\right)^2+A n \left(A (6 n-1) t^2+1\right) 
	\left(A t^2+1\right)^{2 n}\right)}{4 \lambda +8 \pi -1}
\end{equation}
\begin{figure}[H]
	\centering
	(a)\includegraphics[width=5cm,height=4cm,angle=0]{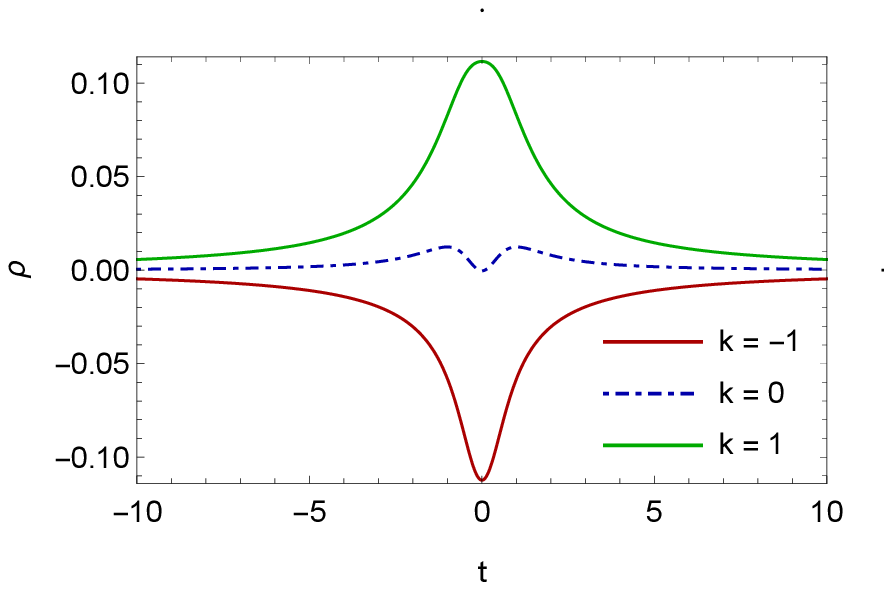}
	(b)\includegraphics[width=5cm,height=4cm,angle=0]{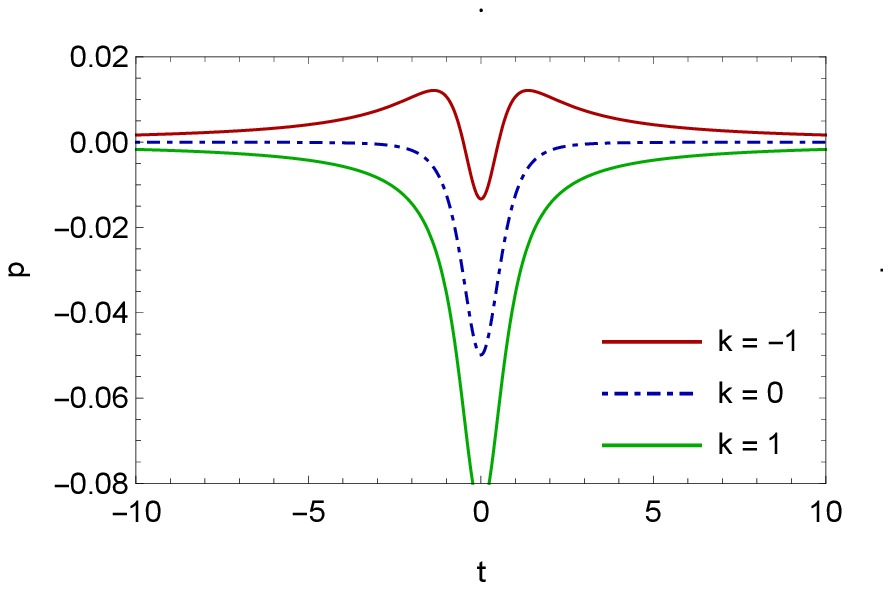}\\
	(c)\includegraphics[width=5cm,height=4cm,angle=0]{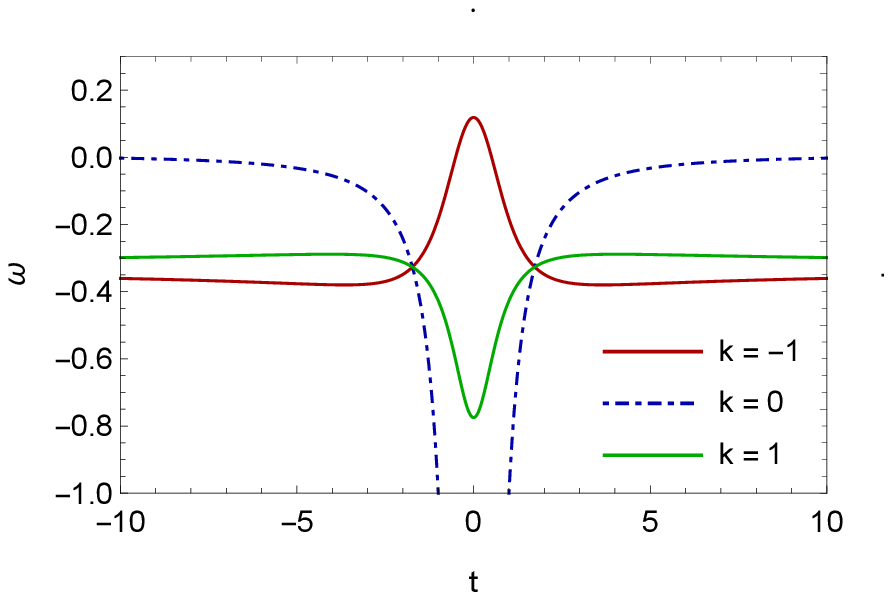}
	(d)\includegraphics[width=5cm,height=4cm,angle=0]{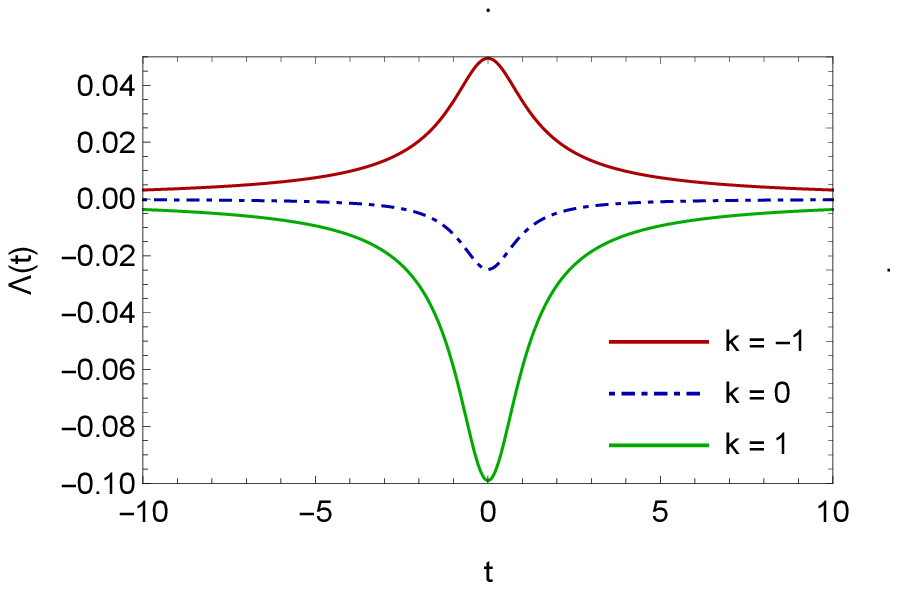}
	\caption{ Evolution of $\rho$, $p$, $\omega$ and $\Lambda$ for the matter bounce scenario ($n=\frac{1}{3}$), $A=1$. (a) The physically 
		accepted behavior of energy density occurs only for a closed universe. (b) The pressure is always negative for the closed universe. 
		(c) The Equation of State parameter for a closed universe lies in the Quintessence region $-1<\omega<0$. (d) The cosmological constant always 
		has a tiny negative value for $k=1$. }\label{figca}	
\end{figure}
Figure (\ref{figca}) shows the evolution of $\rho(t)$, $p(t)$, $\omega(t)$ and $\Lambda(t)$. The evolution of $\rho(t)$ shows that the only case allowed physically is the one with positive curvature $k=+1$. the plots of $p(t)$ and $\omega(t)$ for the closed universe shows a Quintessence-dominated universe along with negative pressure. The varying cosmological constant $\Lambda(t)$ for $k=1$ always has a tiny negative value. While this doesn't seem to agree with some observations which show a tiny positive value of $\Lambda$ ($\approx 10^{-123}$) \cite{84}, the negative value of $\Lambda$ can also fit a large data set and solve the eternal acceleration problem \cite{85}. The negative $\Lambda$ approach has been studied by many authors \cite{85,86,87,88,89,90}. The negative $\Lambda$ is also supported in the framework of the AdS/CFT correspondence \cite{91}. It has been shown in \cite{87} that observationally viable cosmologies with $\Lambda<0$ can be obtained in a modified FRW cosmology. A stable solution with $\Lambda<0$ can also be obtained in Gauss-Bonnet gravity \cite{88}. The interpretation of the varying $\Lambda(t)$ as thermodynamic pressure, which has been suggested in \cite{83}, is itself a natural consequence of negative $\Lambda$ which means a positive pressure of the vacuum. A cyclic cosmology with $\Lambda<0$ has been investigated in \cite{82,92}. The connection between $\Lambda$ and pressure was also studied in \cite{93,94} from different perspectives.\\
{\bf Violation of Energy-Momentum Conservation}
Friedman models in GR ensure the energy conservation through the continuity equation
\begin{equation}\label{14}
\dot{\rho}+3H(\rho+p)=0
\end{equation}
which implies $ d(\rho V ) = -\rho dV  $. Here, $ V = a^3 $ is the volume of universe and $\rho V $ stands for total energy in the universe. With the expansion of universe, dark energy increases in proportion to the volume of expanding universe. If this spacetime still exists then total energy would be constant.\\
Taking a covariant derivative of Eq. (\ref{2}), one can obtain \cite{95,96,97,98}
\begin{equation}\label{15}
\nabla^{i} T_{ij}=\frac{F_{2}(R,T)}{16\pi-F_{2}(R,T)}\left[(T_{ij}+\Theta_{ij})\nabla^{i}\ln F_{2}(R,T)+\nabla^{i}\Theta_{ij}-
\frac{1}{2}g_{ij}\nabla^{i}T\right]
\end{equation}
On substituting $ f(R,T)=R+\lambda T $, Eq.(\ref{15}) reduces to
\begin{equation}\label{16}
\nabla^{i} T_{ij}=-\frac{2 \lambda}{16\pi+2 \lambda}\left[\nabla^{i}(p g_{ij})+\frac{1}{2}g_{ij}\nabla^{i}T\right]
\end{equation}
One should note here that, for $ \lambda=0 $, one would get $ \nabla^{i}T_{ij}=0 $. However for $ \lambda\neq0 $, the conservation of energy-momentum is violated. Recently some researchers have investigated the consequence of the violation of energy-momentum conservation (i.e. $ \dot{\rho}+3H(\rho+p)\neq0 $) in modified gravity theories. It is believed that in phenomenological models, the non-unitary modifications of quantum mechanics with space time discreteness at Planck scale may lead to non-conservation of energy-momentum \cite{99}. Josset et al. \cite{100} have found the non-conservation of energy-momentum leading to an effective cosmological constant which increases or decreases with the creation or annihilation of energy during cosmic expansion and can be reduced to a constant when matter density diminishes. The violation of energy-momentum conservation leading to accelerated expansion of the universe have been observed by Shabani et al. \cite{99} and Sahoo et al. \cite{101} in $f(R,T)$ theories with the pressure less cosmic fluid. We infer from Eq. (\ref{16}) that, for non zero value of parameter $ \lambda $, the conservation of energy-momentum is violated. We quantify the violation of energy-momentum conservation through deviation factor $S$ defined by equation 
\begin{equation}\label{17}
S=\dot{\rho}+3H(\rho+p)
\end{equation}
To satisfy the conservation, $S$ must zero otherwise situation leads to non-conservation. For outward flow of energy from the matter field, $S$ is positive and for inward flow it is negative. The deviation factor $S$ (from equation \ref{15}) must zero for conservation of energy-momentum but, except for too short span of time, conservation is violated in a cosmic cycle.
\section{Physical Properties of the model}
For the present model,the energy conditions(ECs) are expressed as:
\begin{eqnarray}\label{18}
\rho+p&=&\frac{k \left(A t^2+1\right)^{-2 n+2}+2 A n \left(A t^2-1\right)}{(\lambda +4 \pi ) \left(A t^2+1\right)^2}.
\end{eqnarray}
\begin{eqnarray}\label{19}
\rho+3p&=&\frac{\left(A t^2+1\right)^{-2 (n+1)}}{\lambda  (4 \lambda -1)+4 \pi  (6 \lambda -1)+32 \pi ^2}\nonumber\\
&\times&\bigg[2 k (2 \lambda -1) \left(A t^2+1\right)^2-4 A n \left(A t^2+1\right)^{2 n} \big(A t^2 (-5 \lambda +6 \lambda  n+1)\nonumber\\
&+&12 \pi \left(A (2 n-1) t^2+1\right)+5 \lambda -1\big)\bigg].
\end{eqnarray}
\begin{eqnarray}\label{20}
\rho-p&=& \frac{4 \big[k \left(A t^2+1\right)^{-2 n+2}+A n \big(A (6 n-1) t^2+1\big)\big]}{(4\lambda +8 \pi -1) \left(A t^2+1\right)^2}
\end{eqnarray}
\begin{eqnarray}\label{21}
\rho^2-p^2&=&\frac{4 \left(A t^2+1\right)^{-4 (n+1)}}{\lambda  (4 \lambda -1)+4 \pi  (6 \lambda -1)+32 \pi ^2}\nonumber\\
&\times&\bigg[2 A^2 n^2 \left(A t^2-1\right) \left(A (6 n-1) t^2+1\right) \left(A t^2+1\right)^{4 n}+k^2 \left(A t^2+1\right)^4 \nonumber\\
&+&A k n \left(A (6 n+1) t^2-1\right) \left(A t^2+1\right)^{2 n+2}\bigg]
\end{eqnarray}
\begin{eqnarray}\label{22}
\rho p^3&=&-\frac{\left(A t^2+1\right)^{-8 (n+1)}}{16 \left(-4 \lambda ^2-24 \pi  \lambda +\lambda -32 \pi ^2+4 \pi \right)^4} \nonumber\\
&\times& \bigg[ \bigg((1+8 \pi ) k \left(A t^2+1\right)^2+2 A n \left(A t^2+1\right)^{2 n} \big(A t^2 (-6 \lambda +12 \lambda  n+1)\nonumber\\
&+&16 \pi  \left(A (3 n-1) t^2+1\right)+6 \lambda -1\big)\bigg)^3 \bigg[k (8 \lambda +24 \pi -1) \left(A t^2+1\right)^2\nonumber\\
&+&2 A n \left(A t^2+1\right)^{2 n} \left(A t^2 (2 \lambda +12 (\lambda +4 \pi ) n-1)-2 \lambda +1\right)\bigg]\bigg]
\end{eqnarray}
\begin{eqnarray}\label{23}
\rho^2+3p^2&=&\frac{\left(A t^2+1\right)^{-4 (n+1)}}{4 \left(\lambda  (4 \lambda -1)+4 \pi(6 \lambda -1)+32 \pi ^2\right)^2}\nonumber\\
&\times& \bigg[3 \bigg((1+8 \pi ) k \left(A t^2+1\right)^2+2 A n \left(A t^2+1\right)^{2 n} \big(A t^2 (-6 \lambda +12 \lambda  n+1)\nonumber\\
&+&16 \pi  \left(A (3 n-1) t^2+1\right)+6 \lambda -1\big)\bigg)^2+\bigg(k (8 \lambda +24 \pi -1) \left(A t^2+1\right)^2\nonumber\\
&+&2 A n \left(A t^2+1\right)^{2 n} \left(A t^2 (2 \lambda +12 (\lambda +4 \pi ) n-1)-2 \lambda +1\right)\bigg)^2\bigg]
\end{eqnarray}
The speed of sound in the present is found as: 
\begin{eqnarray}\label{24}
C_{s}^{2}&=&\frac{dp}{d\rho}=-\bigg[(1+8 \pi ) k \left(A t^2+1\right)^2+A \left(A t^2+1\right)^{2 n} \bigg(A t^2 (-6 \lambda +12 \lambda  n+1)\nonumber\\
&+&16 \pi \left(3 n \left(A t^2-1\right)-A t^2+3\right)-3 (-6 \lambda +4 \lambda  n+1)\bigg)\bigg]/\bigg[k (8 \lambda +24 \pi -1) \left(A t^2+1\right)^2\nonumber\\
&+&A \left(A t^2+1\right)^{2 n} \left(12 (\lambda +4 \pi ) n \left(A t^2-1\right)+(2 \lambda -1) \left(A t^2-3\right)\right)\bigg]
\end{eqnarray}
The classical linear energy conditions (ECs) \cite{102} ( the null $\rho + p\geq 0$; weak $\rho \geq 0$, $\rho + p\geq 0$; strong $\rho + 3p\geq 0$ and dominant $\rho \geq \left|p\right|$ energy conditions ) need to be replaced by other nonlinear ECs \cite{76,77,78} when semiclassical effects are considered \cite{103}. In addition to the classical ones, we will also consider the following three nonlinear ECs: 
\begin{itemize}
	\item[(i)] The flux EC (FEC): $\rho^2 \geq p_i^2$ \cite{104,105}, first presented in \cite{104}.
	\item[(ii)]  The determinant EC (DETEC): 
	$ \rho . \Pi p_i \geq 0$ \cite{103}.
	\item[(ii)] The trace-of-square EC (TOSEC): $\rho^2 + \sum p_i^2 \geq 0$ \cite{103}.
\end{itemize}
The highly restrictive strong energy condition (SEC) is not expected to be satisfied in the current work because of the negative cosmic pressure (a repulsive gravity) Fig.3(b). The null energy condition (NEC) (Fig. 3(a)) and the dominant energy condition (Fig. 3(c)) are valid all the time only for ($k=+1$). While the violation of the NEC ($\rho + p\geq 0$) occurs in most bouncing models, keeping such condition valid would be highly preferable. It is the most fundamental of all ECs \cite{106} and its violation implies the violation of all other ECs.\\

A classical bouncing model with no violation of the NEC has been presented in \cite{107}. A very useful discussion on the enforcement of the NEC in bouncing cosmology has been given in  \cite{108}.  It has been shown in \cite{109} that non-singular bounces, where the NEC is satisfied, can be obtained by considering a combination of positive curvature and vacuum energy (violating the SEC). Recalling the DE assumption with its negative pressure, the result we have obtained in the present work agrees with the result of the work in \cite{109}. 
We have obtained a combination of positive curvature, violation of the SEC, and a bouncing non-singular universe with no violation of the NEC. The nonlinear ECs have been plotted in Fig. 3(d),(e),(f). Both the flux and trace-of-square ECs are satisfied for $k=+1$.
\begin{figure}[H]
	\centering
	(a)\includegraphics[width=5cm,height=4cm,angle=0]{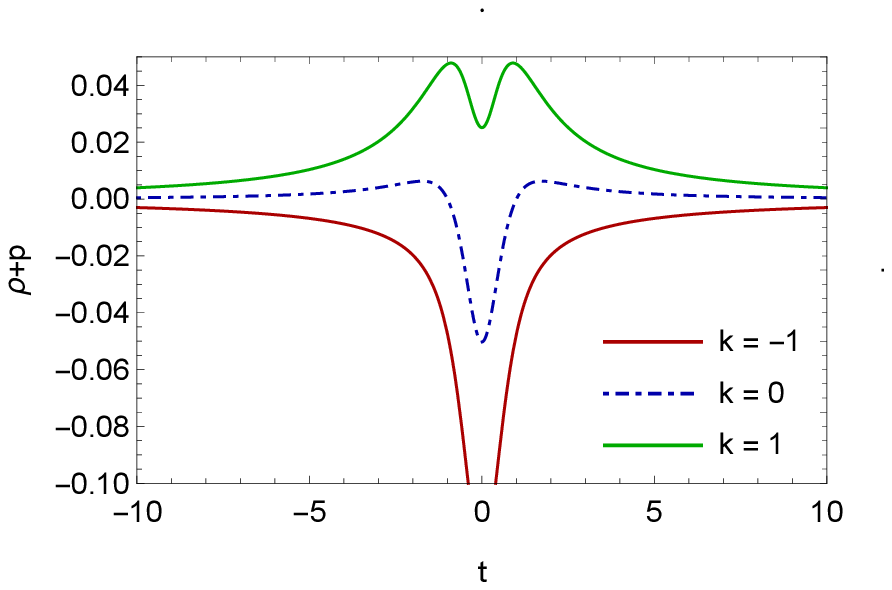}
	(b)\includegraphics[width=5cm,height=4cm,angle=0]{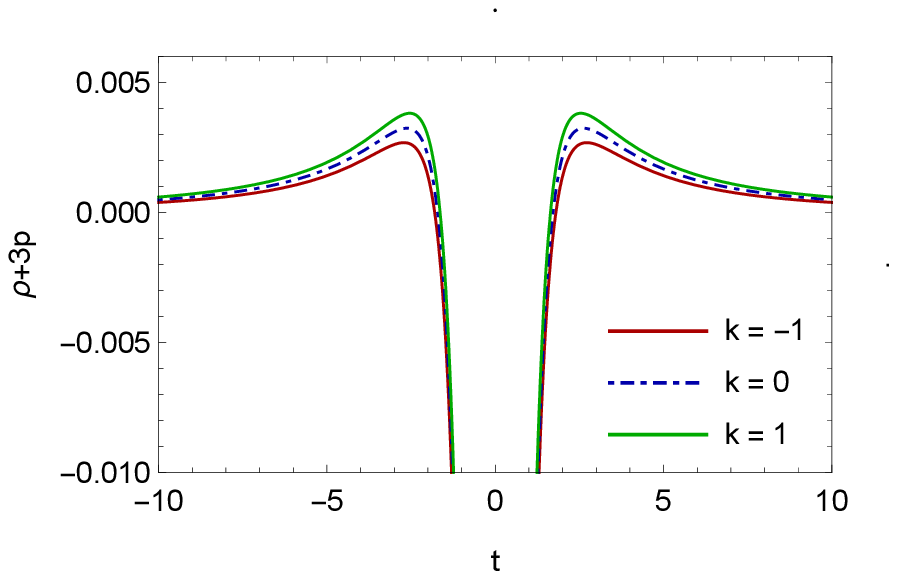}
	(c)\includegraphics[width=5cm,height=4cm,angle=0]{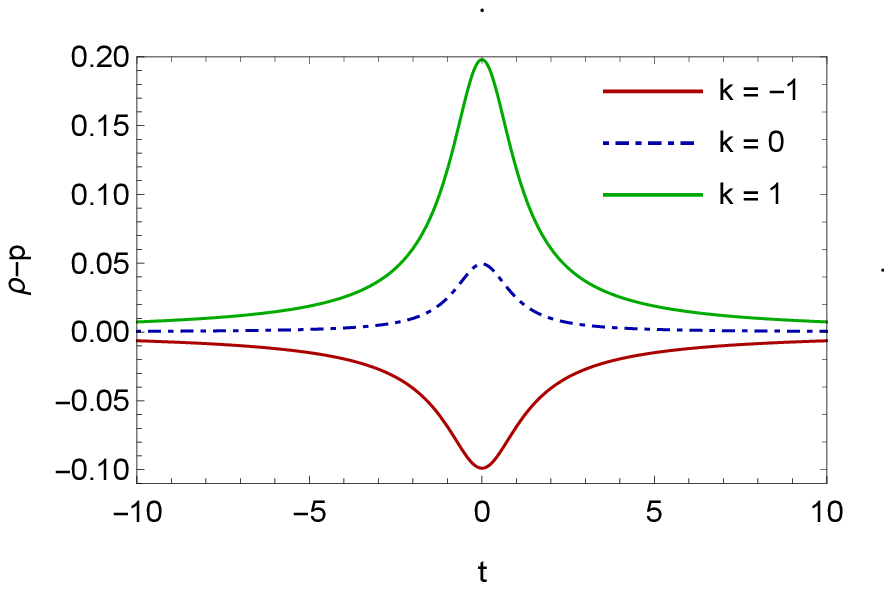}
	(d)\includegraphics[width=5cm,height=4cm,angle=0]{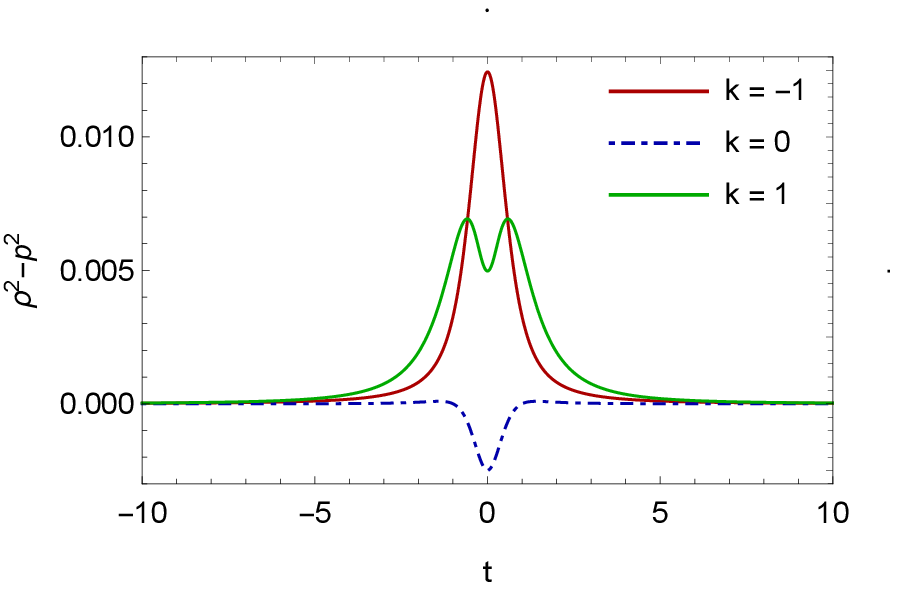}
	(e)\includegraphics[width=5cm,height=4cm,angle=0]{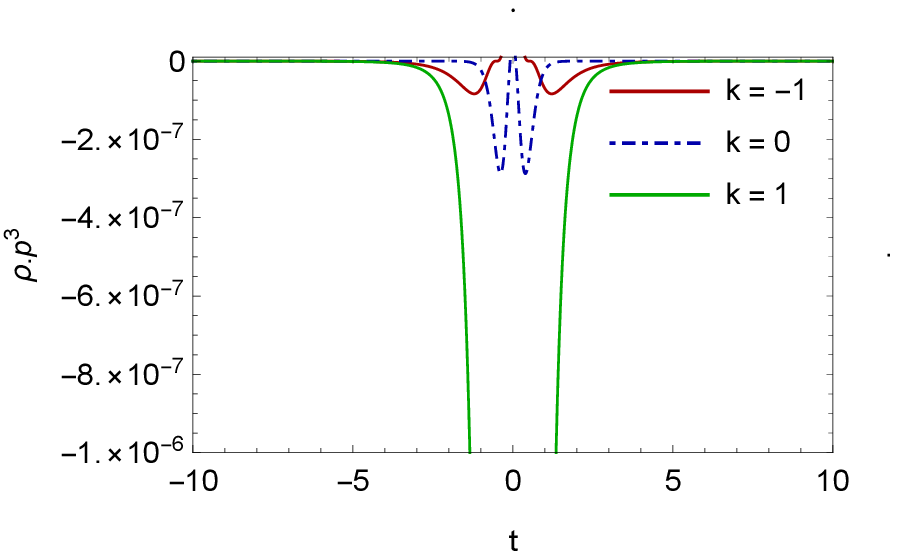}
	(f)\includegraphics[width=5cm,height=4cm,angle=0]{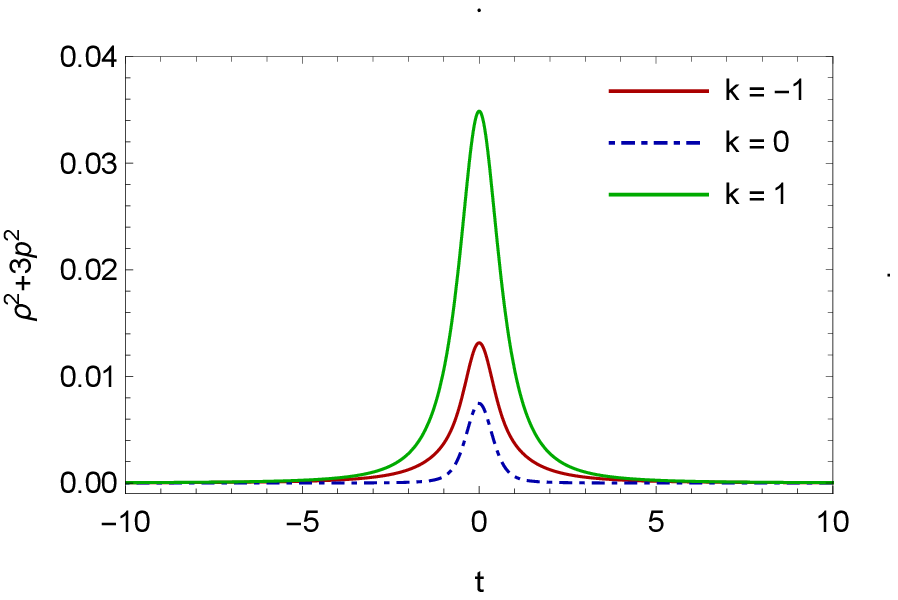}
	(g)\includegraphics[width=5cm,height=4cm,angle=0]{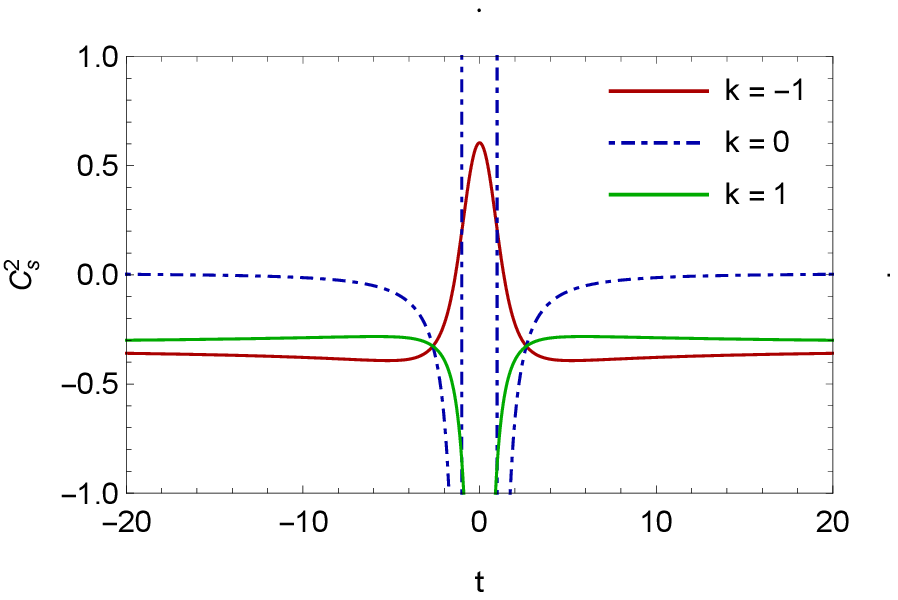}
	\caption{ Classical and nonlinear ECs: No violations of NEC and DEC for $k=+1$. Figure (g) also shows that the causality condition 	$0<\frac{dp}{d \rho}\leq 1$ is satisfied only for the positive curvature except at the bounce.. }\label{fig:cassimir55}	
\end{figure}
\section{Cosmographic Analysis}
For these models, we undertake a cosmographic study in this section.  The significant descriptions of cosmography are given in Refs. \cite{110,111}. As a result, around the present time, we extend scale factor $a(t)$ in Taylor series: 
\begin{equation}\label{25}
a(t) = a_{0}\left[1+\sum_{n=1}^{\infty}\frac{1}{n!}\frac{d^{n}a}{dt^{n}}(t-t_{0})^{n}\right]
\end{equation} 
Equation (\ref{25}), after expanding, provides the most useful terms of cosmographic series, for example, the Hubble parameter $H$, deceleration parameter $q$, jerk $j$, snap $s$ and lerk $l$ parameters
\begin{equation}
\label{26}
H = \frac{1}{a}\frac{da}{dt},
\end{equation}
\begin{equation}
\label{27}
q = -\frac{1}{aH^{2}}\frac{d^{2}a}{dt^{2}},
\end{equation}
\begin{equation}\label{28}
j = \frac{1}{aH^{3}}\frac{d^{3}a}{dt^{3}},
\end{equation}
\begin{equation}\label{29}
s = \frac{1}{aH^{4}}\frac{d^{4}a}{dt^{4}},
\end{equation}
and
\begin{equation}\label{30}
l = \frac{1}{aH^{5}}\frac{d^{5}a}{dt^{5}}.
\end{equation}
For the present model, the expressions for $j$, $l$ and $s$ are obtained as
\begin{equation}\label{31}
j =-\frac{(n-1) [A (2 n-1) t^2+3]}{2 A n^2 t^2}
\end{equation}
\begin{equation}\label{32}
s = -\frac{(n-1) [A^2 (4 n^2-8 n+3) t^4+6 A (2 n-3) t^2+3]}{4 A^2 n^3 t^4}
\end{equation}
\begin{equation}\label{33}
l = -\frac{(n-2) (n-1) [A^2 (4 n^2-8 n+3) t^4+10 A (2 n-3) t^2+15]}{4 A^2 n^4 t^4}
\end{equation}
\begin{figure}[H]
	\centering
	(a)\includegraphics[width=5cm,height=4cm,angle=0]{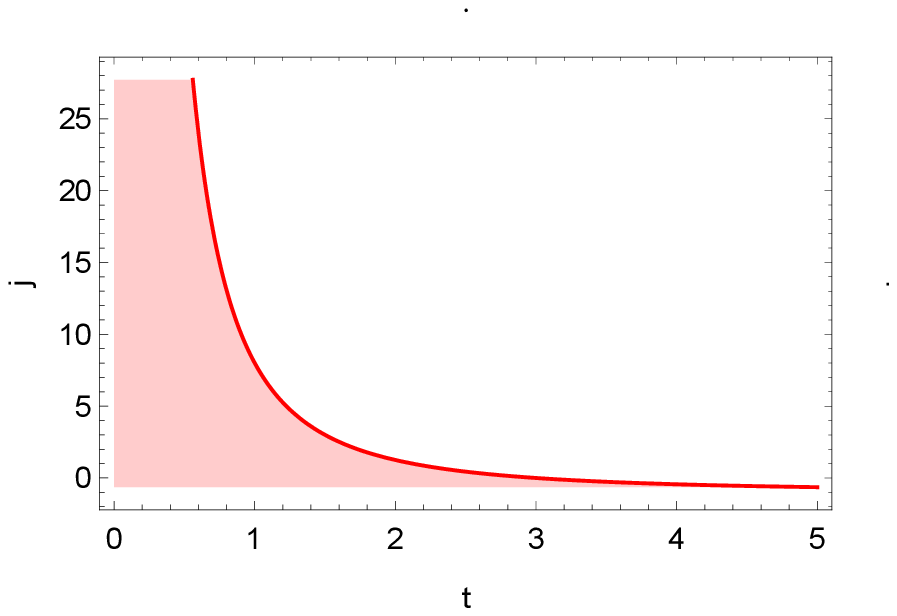}
	(b)\includegraphics[width=5cm,height=4cm,angle=0]{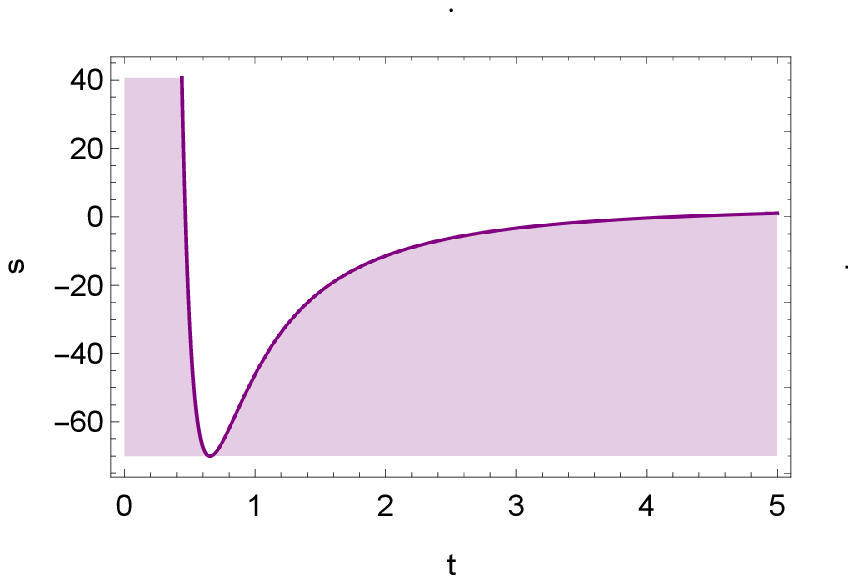}
	(c)\includegraphics[width=5cm,height=4cm,angle=0]{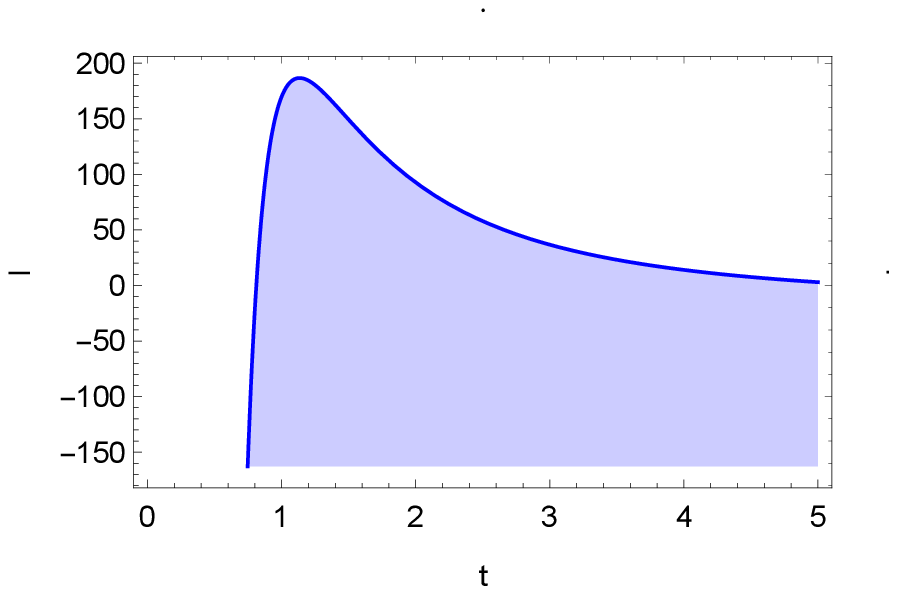}
	\caption{The jerk parameter($j$), snap parameter($s$), and lerk parameter($l$), for the MBS ($n=\frac{1}{3}$) and $A=1$.}\label{fig:cosmog}	
\end{figure}
$H$ and $q$ are useful to study the dynamics of the universe. From the shape of the Hubble curve, it is possible to analyze the physics behind each coefficient and the sign of $q$. This shows whether the dynamics are accelerated or decelerated. $-1<q<0$ represents an expanding universe as expected by current observations. $q =-1$ indicates that the total cosmological energy of universe is dominated by a de Sitter fluid. The other cosmographic terms are jerk $j$ and snap $s$ which are used to discriminate various dark energy models. $j>0$ implies that the universe acceleration started at a precise time during the evolution, associated to the transition redshift. In such a way, it provides the acceleration changes sign during time. In recent review, Capozziello et al. \cite{110} have studied various cosmographic terms and its usefulness in the framework of extended theory of gravity.
\section{Condition of Stability}
The universe was very smooth at early times and it is very lumpy now. cosmologists believe that the reason is gravitational instability. Small fluctuations in the density of the primeval cosmic fluid that grew gravitationally into the galaxies, the clusters and the voids we observe today. Jeans  showed that a homogeneous and isotropic fluid is unstable to small perturbations in its density \cite{112}. The density inhomogeneities grow in time when the pressure support is weak compared to the gravitational pull. As long as pressure is negligible, an over-dense region will keep accrediting material from its surroundings, becoming increasingly unstable until it eventually collapses into a gravitationally bound object. All the structure that we observe around us today originated from minute perturbations in a cosmic fluid that was smooth to the accuracy of one part in ten thousand at the time of recombination. Such tiny irregularities could have been triggered by quantum fluctuations that were stretched out during the inflationary expansion.

To check the stability of the present solution with respect to perturbation of  the metric, the considered perturbations in three expansion 
factors $ a_{i} $ are as \cite{113}
\begin{equation}\label{34}
a_{i}\longrightarrow a_{B_{i}}+\delta a_i=a_{B_{i}}(1+\delta b_i)
\end{equation}
With reference to equation (\ref{34}), the relations representing the perturbation of volume scalar, directional Hubble factors and mean Hubble factor are 
\[ V\longrightarrow V_B+V_B\sum_{i}\delta b_i \]
\begin{equation}\label{35}
\theta_{i}\longrightarrow \theta_{B_{i}}+\sum_{i} \delta b_i
\end{equation}
\[ \theta\longrightarrow \theta_B+\frac{1}{3}\sum_{i}\delta b_i \]
For metric perturbation $ \delta b_i $ to be linear the following equations must be satisfied
\begin{equation}\label{36}
\sum_{i}{\ddot{\delta b_{i}}}+2 \sum_{i} \theta_{B_{i}} \dot{\delta b_{i}}=0
\end{equation}
\begin{equation}\label{37}
\ddot{\delta b_{i}}+\frac{\dot{V_{B}}}{V_{B}}\dot{\delta b_{i}}+\sum_{j} \dot{\delta b_{j}}\theta_{B_{i}}=0
\end{equation}
\begin{equation}\label{38}
\sum \dot{\delta b_{i}}=0
\end{equation}
\begin{figure}[H]
	\centering
	\includegraphics[scale=0.9]{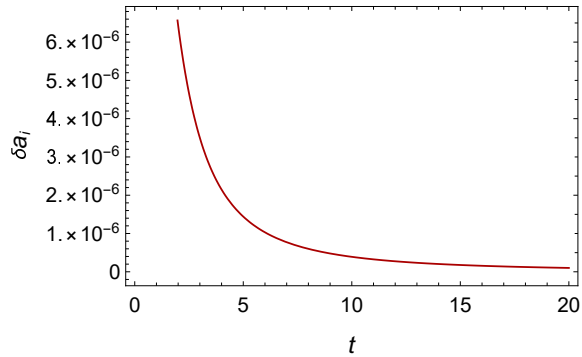}
	\caption{Plot of stability condition.}\label{fig5} 
\end{figure}
From equations (\ref{34})-(\ref{38})
\begin{equation}\label{39}
\ddot{\delta b_{i}}+\frac{\dot{V_B}}{V_B}\dot{\delta b_i}=0.
\end{equation}
where the background volume scalar $ V_B $ leads to 
\begin{equation}\label{40}
V_B=\left(A t^2+1\right)^{3n}
\end{equation}
From equations (\ref{39}) and (\ref{40}), the metric perturbation becomes
\begin{equation}\label{41}
\delta b_i=c_2+t c_1 \, Hypergeometric2F1\left[\frac{1}{2},3 n,\frac{3}{2},-A t^2\right]
\end{equation}
where $ c_1 $ and $ c_2 $ are constant of integration.\\
Thus, the actual fluctuation for each expansion factor $ \delta a_i=a_{B_{i}}\delta b_i $ are expressed as
\begin{equation}\label{42}
\delta a_i=c_2 \left(A t^2+1\right)^{-3n} +c_1 t\left(A t^2+1\right)^{-3n} \, Hypergeometric2F1\left[\frac{1}{2},3 n,\frac{3}{2},-A t^2\right]
\end{equation}
Figure-\ref{fig5}, $\delta a_{i}$ versus cosmic time $ t $, model shows a negligible but significant perturbation initially which decreases 
and soon becomes zero in very short span of time. Thus, in view of large scale measurement, the solution of present problem is stable against 
perturbation of the gravitation field.
\section{Conclusion}
In this paper, we have proposed a matter bouncing model in $ f(R,T) $ gravity. We have considering a special ansatz $ a(t)=\left(A t^2+1\right)^n $ for a variant non-singular bounce \cite{17} and the exact solutions of the field equations in the derived models have been obtained by taking proper physical assumptions. The main features of the present investigation are as follows:
\begin{itemize}
\item We have analyzed the bouncing behavior of the universe in the context of $ f(R,T)$ gravity. The initial circumstances of big-bang cosmology are unknown due to the vanishing scale factor, which leads to singularity. The non-vanishing parameterization of the scale factor, on the other hand, provides the essential initial conditions. $ H $ should be negative before and positive after the bounce for a successful bounce, as shown in figure $1$ (b). 
\item The evolution of $\rho(t)$ shows that the only case allowed physically is the one with positive curvature $k=+1$. the plots of $p(t)$ and $\omega(t)$ for the closed universe show a Quintessence-dominated universe along with negative pressure. The varying cosmological constant $\Lambda(t)$ for $k=1$ always has a 	tiny negative value. The interpretation of the varying $\Lambda(t)$ as thermodynamic pressure is itself a natural consequence of negative $\Lambda$ which means a positive pressure of the vacuum. The dynamical behavior of $\rho(t), $p(t), $\omega(t)$, and $\Lambda(t)$ are depicted in Figure (\ref{figca}). 
\item A bouncing non-singular closed universe without violation of the NEC has been introduced in $f(R, T)$ gravity. The pressure is always $<0$ and the SEC is violated. The model is supported by some recent observations which suggest a closed universe. The new nonlinear ECs have also been investigated. Our result is supported by the work in \cite{109} where the combination of a closed universe and vacuum energy (violating the SEC) results in non-singular bounces where the NEC is satisfied. The nonlinear ECs have been plotted in Fig. $3$(d),(e),(f). Both the flux and trace-of-square ECs are satisfied for $k=+1$. Figure $3$(g) also shows that the causality condition $0<\frac{dp}{d \rho}\leq 1$ is satisfied only for the positive curvature except at the bounce. \item In the cosmographic series of the generated cosmological models, we have assessed different cosmographic parameters like deceleration parameter, Hubble, jerk, snap  and lerk parameters. $-1<q<0$ shows an accelerated expanding universe  as observed by recent observations. For $q =-1$, the total energy of the universe is dominated by a de Sitter fluid. $j>0$ implies that the universe started acceleration during the evolution associated with the transition redshift. In such a way, it provides the acceleration changes sign during the time. 
\item This model, under the assumption of bouncing scenario have shown conclusively, the non-conservation of energy-momentum except for very short duration of cosmic time. The truthfulness of the model has been verified through the stability conditions shown in figure-\ref{fig5}. 
\item In the derived model, the accelerated expansion of the universe have been observed due to violation of conservation of energy-momentum in $f(R,T)$ theory. From Eq. (\ref{16}), we observed that, for non zero value of parameter $\lambda $, the conservation of energy-momentum is violated.
\end{itemize} 
Thus, the bouncing cosmological models suggest a transitioning universe without suffering a singularity.
\section*{Acknowledgement} 
A. Pradhan thanks to the IUCAA, Pune, India for providing facility under associateship programmes. The authors are heartily grateful to the anonymous reviewers for their constructive comments, which improved the paper in the present form.

\end{document}